\documentclass[aps,prl,twocolumn,showpacs,amsmath,amssymb]{revtex4-1}
\usepackage{amsmath}
\usepackage{graphicx}
\usepackage{subfigure}
\usepackage{epstopdf}
\usepackage{color}
\usepackage{multirow}
\usepackage{setspace}
\usepackage{overpic}
\usepackage{amssymb}
\usepackage[bookmarksnumbered, pdfstartview=FitH,colorlinks,urlcolor=blue, citecolor=blue,linkcolor=blue] {hyperref}
\usepackage{lineno}
\usepackage{bm}
\usepackage{rotating}
\usepackage[utf8]{inputenc}

\let\oldequation\equation
\let\oldendequation\endequation

\renewenvironment{equation}
  {\linenomathNonumbers\oldequation}
  {\oldendequation\endlinenomath}

\begin{document}

\title{\bf \boldmath
Measurement of the doubly Cabibbo-suppressed decay $D^+\to K^+\pi^+\pi^-\pi^0$ with semileptonic tags}

\author{
M.~Ablikim$^{1}$, M.~N.~Achasov$^{10,c}$, P.~Adlarson$^{67}$, S. ~Ahmed$^{15}$, M.~Albrecht$^{4}$, R.~Aliberti$^{28}$, A.~Amoroso$^{66A,66C}$, M.~R.~An$^{32}$, Q.~An$^{63,49}$, X.~H.~Bai$^{57}$
% ...
\\
\vspace{0.2cm}
(BESIII Collaboration)\\
\vspace{0.2cm} {\it
$^{1}$ Institute of High Energy Physics, Beijing 100049, People's Republic of China\\
% ...
$^{4}$ Bochum Ruhr-University, D-44780 Bochum, Germany\\
% ...
$^{10}$ G.I. Budker Institute of Nuclear Physics SB RAS (BINP), Novosibirsk 630090, Russia\\
% ...
$^{15}$ Helmholtz Institute Mainz, Staudinger Weg 18, D-55099 Mainz, Germany\\
% ...
$^{28}$ Johannes Gutenberg University of Mainz, Johann-Joachim-Becher-Weg 45, D-55099 Mainz, Germany\\
% ...
$^{32}$ Liaoning Normal University, Dalian 116029, People's Republic of China\\
% ...
$^{49}$ State Key Laboratory of Particle Detection and Electronics, Beijing 100049, Hefei 230026, People's Republic of China\\
% ...
$^{57}$ University of Jinan, Jinan 250022, People's Republic of China\\
% ...
$^{63}$ University of Science and Technology of China, Hefei 230026, People's Republic of China\\
% ...
$^{66}$ University of Turin and INFN, (A)University of Turin, I-10125, Turin, Italy; (B)University of Eastern Piedmont, I-15121, Alessandria, Italy; (C)INFN, I-10125, Turin, Italy\\
$^{67}$ Uppsala University, Box 516, SE-75120 Uppsala, Sweden\\
% ...
$^{c}$ Also at the Novosibirsk State University, Novosibirsk, 630090, Russia\\
% ...
}
}

\begin{abstract}
  Using an $e^+e^-$ annihilation data sample corresponding to an integrated luminosity of
  $2.93\,\rm fb^{-1}$ collected at a center-of-mass energy of 3.773\,GeV with the BESIII detector,
  the doubly Cabibbo-suppressed decay $D^+\to K^+\pi^+\pi^-\pi^0$ is studied with a semileptonic tag method.
  After removing the decays containing narrow intermediate resonances, $D^+\to K^+\eta$, $D^+\to K^+\omega$,
  and $D^+\to K^+\phi$, the branching fraction for the decay $D^+\to K^+\pi^+\pi^-\pi^0$ is determined
  to be $(1.03 \pm 0.12_{\rm stat} \pm 0.06_{\rm syst})\times 10^{-3}$.
The ratio of the branching fraction for $D^+\to K^+\pi^+\pi^-\pi^0$ to its Cabibbo-favored counterpart $D^+\to K^-\pi^+\pi^+\pi^0$ is measured to be $ (1.65\pm0.21)\%$, corresponding to $(5.73\pm0.73)\tan^4\theta_C$, where $\theta_C$ is the Cabibbo mixing angle.
These results are consistent with our previous measurement with hadronic tags but are significantly larger than other doubly Cabibbo-suppressed decays in the charm sector.
\end{abstract}


\maketitle

\oddsidemargin  -0.2cm
\evensidemargin -0.2cm

Studies of hadronic $D$ ($D^0$ or $D^+$) decays are powerful tools for exploring $D^0$-$\bar D^0$ mixing, charge-parity violation and SU(3)-flavor asymmetry breaking~\cite{ref5,theory_1,theory_2,chenghy1,yufs, yufs2}.
Throughout this Letter, charge conjugate modes are implied.
Investigations of doubly Cabibbo-suppressed (DCS) decays of $D$ mesons are an especially intriguing place to explore charm hadron dynamics.
To date, however, only a few DCS $D$ decays have been measured, as summarized in~\cite{pdg2020}.

Naively, the ratio of the branching fraction for a DCS decay relative to its Cabibbo-favored (CF) counterpart is expected to be
about $(0.5-2)\times {\rm tan}^4\theta_C$~\cite{Lipkin,theory_1}, where $\theta_C$ is the Cabibbo mixing angle.
Recently, BESIII reported an observation of the DCS decay $D^+\to K^+\pi^+\pi^-\pi^0$ with hadronic tags,
giving a branching fraction of $(1.14 \pm 0.08_{\rm stat} \pm 0.03_{\rm syst})\times 10^{-3}$ and a DCS/CF branching fraction ratio of $(6.28\pm0.52)\tan^4\theta_C$~\cite{bes3_DCS_Kpipipi0}.
It is important to confirm this anomalously large branching fraction with a new method.
Moreover, comprehensive measurements of DCS $D$ decays, including isospin-related $D^0$ decays,
are crucial to explore the origin of this large DCS to CF branching fraction ratio.
In the measurements of DCS $D^0$ decays using $e^+e^-$ collision data taken at the $\psi(3770)$ resonance peak,
however, the conventional hadronic tag method suffers from
complicated cross feeds between the events of CF $\bar D^0\to {\it tag}$ vs.~DCS $D^0\to {\it signal}$ and those from DCS $D^0\to {\it tag}$ vs.~CF $\bar D^0\to {\it signal}$.
In this Letter, we introduce and utilize a method using semileptonic $\bar D$ decays to tag the DCS $D$ decays.
This new technique helps avoid the aforementioned troubles because the semileptonic $D^0$ decays have no DCS component and the $D^0$-$\bar D^0$ mixing effect is small. As the first step, the reliability of this semileptonic tag method is validated with $D^+\to K^+\pi^+\pi^-\pi^0$ decays whose branching fraction is known to be the largest among the DCS charm decays.  In addition, the semileptonic branching fractions
of $D^+$ mesons are larger than those for the $D^0$.
This is carried out by analyzing a data sample with an integrated luminosity of 2.93\,fb$^{-1}$~\cite{lum_bes3} collected at a center-of-mass energy of $\sqrt s=$ 3.773~GeV  with the BESIII detector.

Details about the design and performance of the BESIII detector are given in Refs.~\cite{BESIII, Ablikim:2019hff}.
Simulated samples produced with the {\sc geant4}-based~\cite{geant4} Monte Carlo (MC) package which
includes the geometric description of the BESIII detector and the
detector response, are used to determine the detection efficiency
and to estimate the backgrounds. The simulation includes the beam
energy spread and initial state radiation in the $e^+e^-$
annihilations modelled with the generator {\sc kkmc}~\cite{kkmc}.
The inclusive MC samples consist of the production of $D\bar{D}$
pairs with consideration of quantum coherence for all neutral $D$
modes, the non-$D\bar{D}$ decays of the $\psi(3770)$, the initial state radiation
production of the $J/\psi$ and $\psi(3686)$ states, and the
continuum processes. The known decay modes are modelled with {\sc
evtgen}~\cite{evtgen} using the branching fractions taken from the
Particle Data Group (PDG)~\cite{pdg2020}, and the remaining unknown decays of the charmonium states are
modeled by {\sc
lundcharm}~\cite{lundcharm}. Final state radiation~(FSR) is incorporated using the {\sc
photos} package~\cite{photos}.

At $\sqrt s=3.773$ GeV, the $D^+D^-$ pair is produced without additional hadrons.
Candidates in which the DCS decay $D^+\to K^+\pi^+\pi^-\pi^0$ and a semileptonic decay tag $D^-\to K^0 e^-\bar \nu_e$ or $D^-\to K^+\pi^- e^-\bar \nu_e$
are both reconstructed, are called double tag (DT) events.
For each of the two semileptonic tags, the branching fraction for $D^+\to K^+\pi^+\pi^-\pi^0$ can be determined by
\begin{equation}\label{equ:br}
{\mathcal B}_{{\rm DCS}} = \frac{N_{{\rm SL},\,{\rm DCS}}} {2\cdot N_{D^{+}D^{-}}\cdot  {\mathcal B}_{\rm SL}\cdot
\epsilon_{{\rm SL},\,{\rm DCS}}\cdot {\mathcal B}_{\rm sub}},
\end{equation}
where $N_{{\rm SL},\,{\rm DCS}}$ is the yield of the signal DT events in the data sample,
$N_{D^+D^-}=(8296\pm31\pm65)\times 10^3$ is the total number of $D^{+}D^{-}$ pairs quoted from the BESIII previous work~\cite{bes3-crsDD},
${\mathcal B}_{\rm SL}$ is the branching fraction for the semileptonic decay quoted from the PDG~\cite{pdg2020},
$\epsilon_{{\rm SL},\,{\rm DCS}}$ is the efficiency of reconstructing the DT events, ${\mathcal B}_{\rm sub}$ is either the product of branching fractions ${\mathcal B}_{\pi^0\to\gamma\gamma}\cdot {\mathcal B}_{K^0\to\pi^+\pi^-}$ or simply ${\mathcal B}_{\pi^0\to\gamma\gamma}$
for the semileptonic tags of $D^-\to K^0 e^-\bar \nu_e$ and $D^-\to K^+\pi^- e^-\bar \nu_e$, respectively.

The signal DT candidates are required to contain exactly six charged tracks and at least two good photons in the final state.
We use the same selection criteria for $K^\pm$, $\pi^\pm$, $e^-$, $K^0_S$, and $\pi^0$ candidates as those used in
Ref.~\cite{bes3_DCS_Kpipipi0,epjc76,cpc40,bes3-Dp-K1ev,bes3-D-b1enu}.
All charged tracks, except for those from $K^0_{S}$ decays, are required to originate from a region within $|\rm{cos\theta}|<0.93$, $V_{xy}<$ 1\,cm and $|V_{z}|<$ 10\,cm.
Here, $\theta$ is the polar angle of the charged track with respect to the detector axis, $V_{xy}$ and $V_{z}$ are the distances of closest approach of the charged track to the interaction point perpendicular to and along the MDC axis, respectively.
Particle identification (PID) of kaons and pions is performed by using combined $dE/dx$ and TOF information.
Charged tracks with a confidence level for the kaon (pion) hypothesis greater than that for the pion (kaon) hypothesis are assigned as kaon (pion) candidates.

Photon candidates are selected using information from the electromagnetic calorimeter (EMC). The shower time is required to be within 700\,ns of the event start time. The shower energy is required to greater than 25 (50)\,MeV if the crystal with the maximum deposited energy in that cluster is in the barrel~(end cap) region~\cite{BESIII}. The opening angle between the shower direction and the extrapolated position on the EMC of closest  charged track must be greater than $10^{\circ}$.
The $\pi^0$ candidates are formed by photon pairs with invariant mass within $(0.115,\,0.150)$\,GeV$/c^{2}$. To improve resolution, a kinematic fit constraining the $\gamma\gamma$
invariant mass to the $\pi^{0}$ known mass given in~\cite{pdg2020} is imposed on the selected photon pair.

To select $D^+\to K^+\pi^+\pi^-\pi^0$ candidates,
the invariant mass of the $\pi^+\pi^-$ pair is required to be outside the interval $(0.478,0.518)$\,GeV/$c^2$ to reject the dominant peaking background from the singly Cabibbo-suppressed decay $D^+\to K^+K_S^0(\to\pi^+\pi^-)\pi^0$. This requirement corresponds to about five standard deviations of the experimental resolution.
The $D^+\to K^+\pi^+\pi^-\pi^0$ signals are identified with two variables: the energy difference
\begin{equation}
\Delta E \equiv E_{D^+} - E_{\rm beam}
\label{eq:deltaE}
\end{equation}
and the beam-constrained (BC) mass
\begin{equation}
M_{\rm BC} \equiv \sqrt{E^{2}_{\rm beam}-|\vec{p}_{D^+}|^{2}}.
\label{eq:mBC}
\end{equation}
Here, $E_{\rm beam}$ is the beam energy,
$\vec{p}_{D^+}$ and $E_{D^+}$ are the momentum and energy of
the $D^+$ candidate in the rest frame of the $e^+e^-$ system, respectively.
If there are multiple candidates for the hadronic side,
only the one with the minimum $|\Delta E|$ is kept.
The correctly reconstructed $D^+$ candidates concentrate around zero in the $\Delta E$ distribution
and around the known $D^+$ mass in the $M_{\rm BC}$ distribution.
The events satisfying $\Delta E\in(-58,45)$\,MeV are kept for further analysis.

Having reconstructed the hadronic $D^+$ decay,
candidates for $D^-\to K^0(\to K^0_S\to \pi^+\pi^-)e^-\bar \nu_{e}$ or $D^-\to K^+\pi^- e^-\bar \nu_{e}$ are selected from the remaining unused tracks.
The charge of the electron candidate is required to be opposite to that of the $D^+$ candidate.
Electron PID uses the combined $dE/dx$, TOF, and EMC information,
with which the combined confidence levels
under the electron, pion, and kaon hypotheses ($CL_e$, $CL_{\pi}$, and $CL_{K}$) are calculated.
Electron candidates are required to satisfy $CL_e>0.001$ and $CL_e/(CL_e+CL_\pi+CL_K)>0.8$. To reduce the background due to mis-identification
between hadrons and electrons, the energy of the electron candidate deposited in the EMC is further required to be greater than 0.8 times its momentum in the MDC. Then, to partially recover the effects of final state radiations and bremsstrahlung (FSR recovery), the four-momenta of photon(s) within $5^\circ$ of the initial electron direction are added to the electron four-momentum measured by the MDC.

The $K^0_S$ candidates must satisfy the following selection criteria.
The two charged tracks are required to satisfy $|V_{z}|<$ 20\,cm but no $V_{xy}$ requirement is imposed.
They are assigned as $\pi^+\pi^-$ without PID requirement.
A secondary vertex fit is applied to the tracks
and candidates with $\chi^2 < 100$ are retained.
The $K^0_S$ candidates are required to have an invariant mass within the interval $(0.486,0.510)~{\rm GeV}/c^2$ and a decay length greater than two standard deviations of the vertex resolution away from the interaction point.
Because the $D^-\to K^+\pi^- e^-\bar \nu_{e}$ decay is dominated by $D^-\to K^*(892)^0 e^-\bar \nu_{e}$,
the invariant mass of $K^+\pi^-$ is restricted to be within the interval $(0.792,0.992)$~GeV/$c^2$ to suppress backgrounds.
The charged kaon and pion are required to satisfy the same PID criteria as for the hadronic side.
Moreover, the kaon candidate is required to have charge opposite to that of the electron.
To suppress potential backgrounds from hadronic decays with a mis-identified electron,
the invariant masses of the $K_S^0 e^-$ and $K^+\pi^- e^-$ combinations, $M_{K_S^0 e^-}$ and $M_{K^+\pi^- e^-}$, are required to
be smaller than 1.8~GeV/$c^2$.
Furthermore, we require that
the maximum energy of any extra photons ($E^{\rm max}_{\rm extra\,\gamma}$) which have not been used in the reconstruction is less than 0.3~GeV and that there is no extra $\pi^0$ candidate ($N_{\rm extra\,\pi^0}$).

The semileptonic $D^-$ decays are identified using a kinematic quantity defined as
\begin{equation}
M^2_{\mathrm{miss}}\equiv E^2_{\mathrm{miss}}-|\vec{p}_{\mathrm{miss}}|^2.
\end{equation}
Here, $E_{\mathrm{miss}}\equiv E_{\mathrm{beam}}-E_{h}-E_{e^-}$ and $\vec{p}_{\mathrm{miss}}\equiv
\vec{p}_{D^-}-\vec{p}_{h}-\vec{p}_{e^-}$ are the missing energy and momentum of the DT event in the $e^+e^-$ center-of-mass system, in
which $E_{h}$ and $\vec{p}_{h}$ are the energy and momentum of $K^0(K^+\pi^-)$,
$E_{e^-}$ and $\vec{p}_{e^-}$ are the energy and momentum of $e^-$, respectively. The
$M^2_{\mathrm{miss}}$ resolution is improved by constraining the $D^+$ energy to the beam energy and $\vec{p}_{D^-} \equiv {-\hat{p}_{D^+}}\cdot\sqrt{E_{\mathrm{beam}}^{2}-M_{D^+}^{2}}$, where $\hat{p}_{D^+}$ is the unit vector in the momentum direction of the $D^+$ and $M_{D^+}$ is the $D^+$ known mass~\cite{pdg2020}.

\begin{figure}[htbp]
  \centering
  \includegraphics[width=0.5\textwidth]{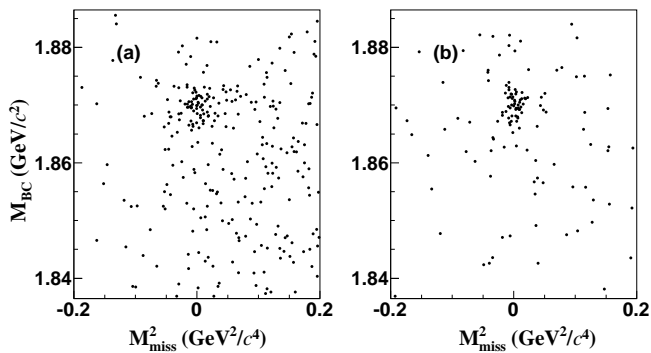}
\caption{Distributions of $M_{\rm BC}$ vs.~$M^2_{\rm miss}$ of the accepted DT candidate events tagged by (a) $D^-\to K^0e^-\bar \nu_e$ and (b) $D^-\to K^+\pi^-e^-\bar \nu_e$ in data. }
\label{fig:M_BC_M2}
\end{figure}

\begin{figure}[htbp]
  \centering
  \includegraphics[width=0.5\textwidth]{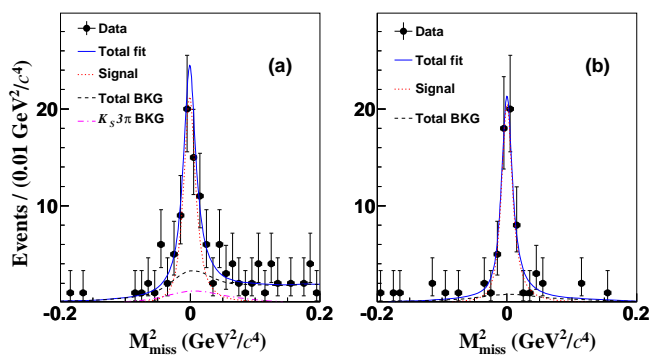}
\caption{Simultaneous fits to the $M^2_{\rm miss}$ distributions of the accepted DT candidate events tagged by (a) $D^-\to K^0e^-\bar \nu_e$ and (b) $D^-\to K^+\pi^-e^-\bar \nu_e$.
Points with error bars are data.
Blue solid curves are the total fit results. Red dotted and black dashed curves are the fitted signal and background (BKG) distributions, respectively.
In (a), the pink dot-dashed curves is the $K_S 3\pi$ BKG
of $D^0\to K^0_S\pi^+\pi^-\pi^0$ vs.~$\bar D^0\to K^+e^-\bar \nu_e$.
}
\label{fig:fits_umiss}
\end{figure}

Figure \ref{fig:M_BC_M2} shows the distributions of $M_{\rm BC}$ vs.~$M^2_{\rm miss}$ for the DT candidates in data. The clusters around the $D^+$ known mass and zero indicate signal DT candidate events.
The candidates satisfying $M_{\rm BC}\in (1.864,1.874)$\,GeV/$c^2$ are kept for further analysis.
With this requirement imposed, the $M^2_{\mathrm{miss}}$ distributions of the survived events are shown in Fig.~\ref{fig:fits_umiss}.

The signal MC events of $D^+\to K^+\pi^+\pi^-\pi^0$ are simulated using the MC generator which was adopted in the previous BESIII work~\cite{bes3_DCS_Kpipipi0}. The generator incorporates the resonant decays $D^+\to K^*(892)^0\rho(770)^+$, $K^*(892)^+\rho(770)^0$, and non-resonant $D^+\to K^+\pi^+\pi^-\pi^0$, including interference effects.  The small contributions from $K^+\eta$, $K^+\omega$, and $K^+\phi$ are then added without considering interference.
The detection efficiencies $\epsilon_{{\rm SL},\,{\rm DCS}}$ are obtained to be $0.103\pm0.001$ and $0.076\pm0.001$ for the DT events tagged
by $D^-\to K^0e^-\bar \nu_e$ and $D^-\to K^+\pi^-e^-\bar \nu_e$, respectively,
where the efficiencies do not include the branching fractions for $K^0$ and $\pi^0$ decays,
and the uncertainties are statistical only.

To extract the signal yield, unbinned maximum likelihood simultaneous fits are performed
on the $M^2_{\mathrm{miss}}$ distributions for the two semileptonic tags.
The dominant (``$K_S 3\pi$'') background of $D^0\to K^0_S\pi^+\pi^-\pi^0$ vs.~$\bar D^0\to K^+e^-\bar \nu_e$
happens for the tag of $D^-\to K_S^0 e^-\bar \nu_e$ due to wrongly positioning $K^0_S$ and $K^+$, 
as shown in Fig.~\ref{fig:fits_umiss}(a). Miscellaneous backgrounds include the decay $D^+\to K_S^0\pi^+\pi^0(\pi^0)$ vs.~$D^-\to K^+\pi^-e^-\bar \nu_e$
faking the signal for the tag of $D^-\to K_S^0e^-\bar \nu_e$ owing to switching $K_S^0$ and $K^+\pi^-$, and
the decay $D^+\to K^+K^-(\to \pi^-\pi^0)\pi^+$ passing the event selections for both the tags. 
The remaining backgrounds comprise the combinatorial background and a small contribution from mis-reconstructed semileptonic candidates.
These backgrounds and the semipletonic signal are described by corresponding MC-simulated shapes derived from the inclusive MC sample.
The yield of the $K_S 3\pi$ background is fixed based on the known branching fractions and the mis-identification rates, and the yields of the signal and
non-$K_S 3\pi$ backgrounds are free parameters of the fits.
The two semileptonic tags are constrained to have the same branching fraction for $D^+\to K^+\pi^+\pi^-\pi^0$.
The fit results are shown in Fig.~\ref{fig:fits_umiss}.
The fits give a total yield of $112\pm12$ for signal DT events, where the uncertainty is statistical only.
This leads to ${\mathcal B}(D^+\to K^+\pi^+\pi^-\pi^0)=(1.11\pm0.12)\times 10^{-3}$, where the uncertainty is statistical only.
The statistical significance of the signal, calculated by $\sqrt{-2{\rm ln ({\mathcal L_0}/{\mathcal
      L_{\rm max}}})}$, is found to be greater than $10\sigma$. Here,
${\mathcal L}_{\rm max}$ and ${\mathcal L}_0$ are the maximal
likelihoods of the fits with and without inclusion of a signal contribution, respectively.

The systematic uncertainties in the branching fraction measurement are estimated relative to the measured branching fraction and discussed below.
The systematic uncertainties originating from $e^-$ tracking (PID) efficiencies
are studied by using the control samples of $e^+e^-\to\gamma e^+ e^-$ events and
those for $K^+$ and $\pi^\pm$ are investigated with partially-reconstructed hadronic $D\bar D$ events.
The efficiency ratios of data and MC simulation for
$e^-$ tracking, $e^-$ PID, $K^+$ tracking, $K^+$ PID, $\pi^\pm$ tracking, and $\pi^\pm$ PID
are
$(100.0\pm0.5)\%$, $(101.2\pm0.2)\%$,
$(102.0\pm0.3)\%$, $(100.0\pm0.2)\%$,
$(100.0\pm0.2)\%$, and $(100.0\pm0.2)\%$, respectively.
Here, the two dimensional (momentum and $\cos\theta$) $e^-$ tracking (PID) efficiencies
from $e^+e^-\to\gamma e^+ e^-$ events and the momentum dependent $K^+(\pi^\pm)$ tracking (PID) efficiencies from the partially-reconstructed hadronic $D\bar D$  events  are re-weighted to match those in the signal decays.
The signal MC efficiencies are corrected by the aforementioned differences where necessary.
After these corrections, the quoted uncertainties on the tracking (PID) efficiency ratios are taken as systematic uncertainties.
The systematic uncertainty related to the $K_{S}^{0}$ reconstruction efficiency
is estimated with the control samples of $J/\psi\to K^{*}(892)^{\mp}K^{\pm}$ and $J/\psi\to \phi K_S^{0}K^{\pm}\pi^{\mp}$~\cite{sysks}.
The associated systematic uncertainty is assigned as 1.6\% per $K^0_S$.
The systematic uncertainty of the $\pi^0$ reconstruction efficiency is investigated by using the partially-reconstructed hadronic $D\bar D$  decays of
$\bar D^0\to K^+\pi^-\pi^0$ and $\bar D^0\to K^0_S\pi^0$ decays tagged by either $D^0\to K^-\pi^+$ or $D^0\to K^-\pi^+\pi^+\pi^-$~\cite{epjc76,cpc40}.
The data to MC efficiency ratio is $(99.7 \pm 0.8)\%$ giving a systematic uncertainty of 0.8\% per $\pi^0$ after the small correction is applied.
The combined effect on the measured branching fraction due to the systematic uncertainties of tracking and PID efficiencies of $K^+$, $\pi^\pm$, and $e^-$
as well as the reconstruction efficiencies of $K^0_S$ and $\pi^0$ is 1.5\%.

The systematic uncertainty in the $M^2_{\rm miss}$ fit is estimated by comparing the nominal branching fraction with the one
measured with alternative signal shapes and background shapes.
The systematic uncertainty due to the signal shape is examined by replacing the nominal shape with one convolved with a Gaussian function.
Its parameters represent the data-MC simulation difference and are
obtained from the CF decay $D^+\to K^-\pi^+\pi^+\pi^0$. The change of branching fraction due to the signal shape is found to be negligible.
The systematic uncertainty from the simulated background shape is taken into account by varying the $K_S 3\pi$ background component by
the uncertainty of the input BF~\cite{pdg2020}, resulting a 1.5\% change of the re-measured branching fraction.
The background contributed by $D^+\to K_S^0\pi^+\pi^0(\pi^0)$ vs.~$D^-\to K^+\pi^-e^-\bar \nu_e$ is tested by varying the input BF by
its listed uncertainty~\cite{pdg2020}. It is conservative to assigned a 0.5\% as a systematic uncertainty.
The influence of the smooth parameters is also examined by
varying the smooth parameters of background shapes and the maximum change of the re-measured branching fraction is 3.5\%. The 
quadratical sum of these three changes, 3.8\%, is assigned as the associated systematic uncertainty.

The systematic uncertainties related to the requirements of $\Delta E$ and $M_{\rm BC}$ for the hadronic side as well
as the requirements of $M_{K^+\pi^-}$, $M_{K^0_Se^-}$, and $M_{K^+\pi^-e^-}$ for the semileptonic sides are studied
by using the control samples of the CF decay $D^+\to K^-\pi^+\pi^+\pi^0$ vs.~the same semileptonic tags in this analysis.
The corresponding uncertainties are taken to be the differences of the acceptance efficiencies between data and MC simulation.
The systematic uncertainty of the $K^0_S$ veto is assigned as the difference of the DT efficiencies
with the $K^0_S$ veto mass windows set with the mass resolutions from data and MC simulation.
These uncertainties are all found to be negligible.

The uncertainty from FSR recovery is assigned to be 0.3\% based on a large sample of $D^0 \to \bar K^- e^+\nu_e$ decays~\cite{bes3-D0-kev}.
The uncertainty due to the limited MC simulation sample size, 0.8\%, is taken into account as a systematic uncertainty.
The systematic uncertainty in the MC modeling of the DCS decay $D^+\to K^+\pi^+\pi^-\pi^0$ is assigned as 1.3\%, which is quoted from Ref.~\cite{bes3_DCS_Kpipipi0}.
In contrast, the associated uncertainties in the MC modeling of the semileptonic decays of $D^-\to \bar K^0e^-\bar \nu_e$ and $D^-\to K^+\pi^- e^-\bar\nu_e$
are negligible~\cite{bes3-Dp-k0ev,bes3-Dp-kpiev}.
The systematic uncertainty arising from the requirements of $E^{\rm max}_{\rm extra\,\gamma}$ and $N_{\rm extra\,\pi^0}$
is estimated by using the control samples of the CF decay $D^+\to K^-\pi^+\pi^+\pi^0$ vs.~the same semileptonic tags in this analysis.
The difference between the data and MC efficiencies, 0.3\%, is assigned as a systematic uncertainty.

The total number of the $D^{+}D^{-}$ pairs in the data sample is quoted from Ref.~\cite{bes3-crsDD} and its uncertainty of 0.9\% contributes a systematic uncertainty.
The branching fractions for $D^-\to K^0e^-\bar \nu_e$ and $D^-\to K^+\pi^-e^-\bar \nu_e$ are quoted from the PDG~\cite{pdg2020}.
Their uncertainties are 1.1\% and 4.4\% and their consequent impact on the measured branching fraction is 2.3\%.

Assuming that all these uncertainties are independent, we determine the total systematic uncertainty to be 5.0\% by adding the above effects quadratically.
The systematic uncertainties discussed above are summarized in Table~\ref{tab:relsysuncertainties}.

\begin{table}[htp]
\centering
\caption{
Systematic uncertainties in the branching fraction measurement.}
\label{tab:relsysuncertainties}
\centering
\begin{tabular}{c|c}
  \hline
  \hline
Source                                    & Uncertainty (\%)\\ \hline
Tracking, PID, $K^0_S$ and $\pi^0$        & 1.5\\
$M_{\rm miss}^{2}$ fit                    & 3.8\\
$\Delta E$ and $M_{\rm BC}$ requirements  & Negligible\\
$K^+\pi^-$ mass window                    & Negligible\\
$K_{S}^{0}$ veto                          & Negligible\\
FSR recovery                             & 0.3\\
MC statistics                             & 0.8\\
MC model                                  & 1.3\\
$E^{\rm max}_{\rm extra\,\gamma}$ and $N_{\rm extra\,\pi^0}$         & 0.3\\
$N_{D^+ D^-}$                             & 0.9\\
Quoted branching fractions                & 2.3\\ \hline
Total                                     & 5.0\\
\hline
\end{tabular}
\end{table}

In conclusion, we introduce a new semileptonic tagging method to investigate DCS $D$ decays and employ it to perform an analysis using $2.93\,\rm fb^{-1}$ of $e^+e^-$ collision data collected at $\sqrt s=3.773$\,GeV with the BESIII detector.
The feasibility of this method is now verified by the independent measurement of the DCS decay $D^+\to K^+\pi^+\pi^-\pi^0$.
After subtracting the sum of the product branching fractions for decays containing narrow intermediate resonances, $D^+\to K^+X\,(X=\eta,\omega,\phi)$ with $X\to \pi^+\pi^-\pi^0$~\cite{note}
and ignoring the possible interference between these decays and the other processes in $D^+\to K^+\pi^+\pi^-\pi^0$,
the branching fraction for $D^+\to K^+\pi^+\pi^-\pi^0$ is measured to be $(1.03\pm0.12\pm0.06)\times 10^{-3}$.
Using the world average value of ${\mathcal B}(D^+\to K^-\pi^+\pi^+\pi^0)$,
we obtain the branching fraction ratio ${\mathcal B}(D^+\to K^+\pi^+\pi^-\pi^0)/{\mathcal B}(D^+\to K^-\pi^+\pi^+\pi^0)=(1.65\pm0.21)\%$,
corresponding to $(5.73\pm0.73)\tan^4\theta_C$.
This confirms the anomalously large rate of $D^+\to K^+\pi^+\pi^-\pi^0$ observed in our previous work~\cite{bes3_DCS_Kpipipi0}.
Because the ratio of the PDG value of ${\mathcal B}(D^0\to K^+\pi^+\pi^-\pi^+)$ over its CF counterpart supports naive expectations, the obtained large ratio is likely caused by differing resonance structures and final state interactions in $D^+\to K^+\pi^+\pi^-\pi^0$. Future amplitude analysis of this decay with a larger data sample will help discover the origin of this unexpected result.
The semileptonic tag method is verified to work well and gives similar precision to our earlier measurements with hadronic tags. It also provides a new technique to access the DCS $D^0$ decays (which are difficult with the traditional hadronic tags) with a larger $\psi(3770)$ data sample~\cite{bes3-white-paper} in the near future.
The results in this work and Ref.~\cite{bes3_DCS_Kpipipi0} demonstrate that
at least some DCS $D$ decays are significantly enhanced.
Further studies in the BESIII~\cite{bes3-white-paper}, Belle II~\cite{belle2-white-paper}, and LHCb~\cite{lhcb-white-paper} experiments are anticipated and will be crucial for further understanding of hadronic charm physics.

The BESIII collaboration thanks the staff of BEPCII and the IHEP computing center for their strong support. This work is supported in part by National Key Research and Development Program of China under Contracts Nos. 2020YFA0406400, 2020YFA0406300; National Natural Science Foundation of China (NSFC) under Contracts Nos. 11775230, 11875054, 11625523, 11635010, 11735014, 11822506, 11835012, 11935015, 11935016, 11935018, 11961141012; the Chinese Academy of Sciences (CAS) Large-Scale Scientific Facility Program; Joint Large-Scale Scientific Facility Funds of the NSFC and CAS under Contracts Nos. U2032104, U1732263, U1832207; CAS Key Research Program of Frontier Sciences under Contracts Nos. QYZDJ-SSW-SLH003, QYZDJ-SSW-SLH040; 100 Talents Program of CAS; INPAC and Shanghai Key Laboratory for Particle Physics and Cosmology; ERC under Contract No. 758462; European Union Horizon 2020 research and innovation programme under Contract No. Marie Sklodowska-Curie grant agreement No 894790; German Research Foundation DFG under Contracts Nos. 443159800, Collaborative Research Center CRC 1044, FOR 2359, FOR 2359, GRK 214; Istituto Nazionale di Fisica Nucleare, Italy; Ministry of Development of Turkey under Contract No. DPT2006K-120470; National Science and Technology fund; Olle Engkvist Foundation under Contract No. 200-0605; STFC (United Kingdom); The Knut and Alice Wallenberg Foundation (Sweden) under Contract No. 2016.0157; The Royal Society, UK under Contracts Nos. DH140054, DH160214; The Swedish Research Council; U. S. Department of Energy under Contracts Nos. DE-FG02-05ER41374, DE-SC-0012069.


\begin{thebibliography}{**}
\bibitem{note}
The branching fraction for $D^+\to K^+\omega$ is taken from Ref.~\cite{bes3_DCS_Kpipipi0}
and those for the other decays are quoted from the PDG~\cite{pdg2020}.
\end{thebibliography}
\end{document}